\documentclass[prl,aps,twocolumn,showpacs,superscriptaddress,dvips,linenumbers]{revtex4}
\usepackage{amsfonts}
\usepackage{amsmath}
\usepackage{amssymb}
\usepackage{graphicx}
\usepackage{graphics}
\usepackage{dcolumn}
\usepackage{bm}

\begin{document}

\title{Asymptotic expressions for the hyperfine populations in the ground
state of spin-1 condensates against a magnetic field }
\affiliation{Shaoguan University, Shaoguan, 512005, P. R. China}
\affiliation{The State Key Laboratory of Optoelectronic Materials and Technologies,
School of Physics and Engineering, Sun Yat-Sen University, Guangzhou, P. R.
China}
\affiliation{State Key Laboratory of Theoretical Physics, Institute of Theoretical
Physics, Chinese Academy of Sciences, Beijing, 100190, China}
\author{Y. M. Liu$^{1,3}$,Y. Z. He$^2$, and C. G. Bao$^{2}$ }
\thanks{Corresponding author: stsbcg@mail.sysu.edu.cn}

\begin{abstract}
Based on the perturbation theory up to the second order, analytical
asymptotic expressions for the variation of the population of hyperfine
component $\mu=0 $ particles in the ground state of spin-1 condensates
against a magnetic field $B$ has been derived. The ranges of $B$ in which
the asymptotic expressions are applicable have been clarified via a
comparison of the numerical results from the analytical expressions and from
a diagonalization of the Hamiltonian in a complete spin-space. It was found
that, For Rb, the two analytical expressions, one for a weak and the other
one for a strong field, together cover the whole range of $B$ from 0 to
infinite. For Na, the analytical expressions are valid only if $B$ is very
weak or sufficiently strong.
\end{abstract}

\maketitle

\subsubsection{1, Introduction}

The spinor condensates, as tunable systems with active spin-degrees of
freedom, are rich in physics and promising in application. Since the
pioneering experiment on spin-1 condensates \cite{ste} the study of these
systems becomes a hot topic.\cite{yka,sta} In general, in the study of spin-$%
f$ condensates, an important observable is the population of the particles
lying in a given hyperfine-component $\mu =-f$ to $f$. These quantities are
popularly measured in various experiments and are a key to relate
experimental results to theories.\cite{ste,dms,hsc,msc,tku} Hence, the
theoretical study of these populations is desirable.

Let the hyperfine density of $\mu =0$ component in the ground state (g.s.)
of spin-1 condensates be denoted as $\bar{\rho}_{0}$. When a magnetic field $%
B$ is applied, the variation of $\bar{\rho}_{0}$\ as a function of $B$\ is
studied in this paper. A many-body theory, instead of the mean-field theory,
is used so that the spin-degrees of freedom are treated rigorously. The
emphasis is placed on the asymptotic behaviors of $\bar{\rho}_{0}$\ when $%
B\rightarrow 0$ and $\rightarrow \infty $. Based on the second order
perturbation theory analytical asymptotic expressions for $\bar{\rho}_{0}$
are derived. Numerical results from a diagonalization of the Hamiltonian are
also obtained to compare with the results from the asymptotic expressions.
In this way the effective ranges of $B$\ wherein the asymptotic expressions
are applicable are clarified. Thus, within the effective ranges, $\bar{\rho}%
_{0}$ can be quantitatively and accurately known, and in general the
variation of $\bar{\rho}_{0}$\ as a function of $B$ can be qualitatively
understood.

\subsubsection{2, Hamiltonian in the spin-space}

We consider $N$ spin-1 cold atoms trapped by an isotropic potential and
subjected to a magnetic field. For spin-1 atoms the dipole-dipole
interaction is extremely weak, and thus its effect on the coupling between
the spatial- and spin-modes can be neglected (the quasi-spin-orbit coupling
is not considered here). Let the total interaction $\hat{V}=\hat{V}_{0}+\hat{%
V}_{2}$, $\hat{V}_{0}=c_{0}\sum_{i<j}\delta (\mathbf{r}_{i}-\mathbf{r}_{j})$
and $\hat{V}_{2}=c_{2}\sum_{i<j}\delta (\mathbf{r}_{i}-\mathbf{r}_{j})\
\mathbf{f}_{i}\cdot \mathbf{f}_{j}$ where $\mathbf{f}_{i}$ is the
spin-operator of the $i$-th particle.\cite{tlh,ckl} It was found that $%
|c_{2}/c_{0}|=0.0046$\ for $^{87}$Rb and 0.031 for $^{23}$Na. Thus\ the
spin-dependent force is nearly two order weaker than the central
(spin-independent) force. Therefore, the spin-modes are much easier to get
excited. Accordingly, the lowest lying levels would avoid spatial
excitations but be dominated by spin-modes. With this in mind, it is assumed
that the g.s. does not contain spatially excited modes, and all the
particles fall into the same spatial state which is most advantageous to
binding. This is the basic assumption of this paper.

The Hamiltonian $\hat{H}$ is%
\begin{equation}
\hat{H}=\hat{H}_{0}+\hat{V}+q\Sigma _{i}f_{iz}^{2}  \label{1}
\end{equation}%
where $\hat{H}_{0}=\sum_{i}(-\frac{\hbar ^{2}}{2m}\nabla _{i}^{2}+U(\mathbf{r%
}_{i}))$ includes the kinetic and trap energies, Since the Hamiltonian
conserves the total magnetization $M$, the linear Zeeman term has been
neglected. The third term\ at the right of eq.(1) is for the quadratic
Zeeman energy, where $q$ is proportional to $B^{2}$.

When $B=0$, not only $M$, the total spin $S$ is also conserved. It has been
proved that, for a spin-1 $N-body$ system, the all-symmetric spin-state with
a pair of good quantum numbers $(S,M)$ is unique, where $N-S$ must be even.%
\cite{ka} Let the total spin-state be denoted as $\vartheta _{SM}$. The set $%
\{\vartheta _{SM}\}$ form a complete set for all-symmetric spin-states.
Thus, under the above basic assumption, the g.s. can be written in the form
as\cite{bao3}
\begin{equation}
\Phi _{\mathrm{gs}}=\Pi _{i}\phi (\mathbf{r}_{i})\vartheta _{S_{g}M}
\label{2}
\end{equation}%
where $\phi (\mathbf{r}_{i})$ is the common spatial state for all the
particles, $S_{g}=N$ for Rb, and $S_{g}=M$ or ($M+1$)\ for Na if $N-M$ is
even or (odd). In what follows $M\geq 0$ and $N-M$ being even are assumed
for convenience.

When $B\neq 0$, $S$ is not conserved but $M$\ is. Hence, different $%
\vartheta _{SM}$ having the same $M$ but distinct in $S$ are mixed up, and
the g.s. can be in general written as
\begin{equation}
\Phi _{\mathrm{gs}}=\Pi _{i}\phi (\mathbf{r}_{i})\sum_{S}C_{S}\vartheta
_{SM}\equiv \Pi _{i}\phi (\mathbf{r}_{i})\Theta _{gs}  \label{3}
\end{equation}

Since the spatial degrees of freedom are considered as being frozen, the Schr%
\"{o}dinger equation can be projected into the spin-space as
\begin{equation}
\int d\mathfrak{R}\Pi _{i}\phi ^{\ast }(\mathbf{r}_{i})(H-E)\Psi =0
\label{4}
\end{equation}%
where the integration covers all the spatial degrees of freedom. Making use
of the fact that $\sum_{i<j}\mathbf{f}_{i}\cdot \mathbf{f}_{j}=\frac{1}{2}%
\overset{\wedge }{S}^{2}-N$, where $\overset{\wedge }{S}$\ is the operator
of the total spin, the above equation becomes
\begin{equation}
(H^{\prime }-E^{\prime })\Theta _{gs}=0  \label{4p}
\end{equation}%
Where $H^{\prime }=H_{0}^{\prime }+H_{1}^{\prime }$, $H_{0}^{\prime }=\frac{1%
}{2}Xc_{2}\overset{\wedge }{S}^{2}$, $H_{1}^{\prime }=q\Sigma _{i}f_{iz}^{2}$%
, $E^{\prime }=E-N\overset{\_}{h}-\frac{N(N-1)}{2}Xc_{0}+NXc_{2}$, $X\equiv
\int d\mathbf{r}|\phi |^{4}$, and $\overset{\_}{h}=\langle \phi |-\frac{%
\hbar ^{2}}{2m}\nabla _{i}^{2}+U(\mathbf{r}_{i})|\phi \rangle $. In what
follows we work simply in the spin-space with the Hamiltonian $H^{\prime }$.

\subsubsection{3, Asymptotic behavior of the hyperfine density when $%
B\rightarrow 0$}

When $B\rightarrow 0$,\ $H_{1}^{\prime }$ can be treated as a perturbation.
Since $\vartheta _{SM}$\ is an eigenstate of $H_{0}^{\prime }$ and the set $%
\{\vartheta _{SM}\}$\ is complete, the perturbative states can be expanded
in terms of $\{\vartheta _{SM}\}$. A crucial point is the calculation of the
matrix element $\langle \vartheta _{S^{\prime }M}|H_{1}^{\prime }|\vartheta
_{SM}\rangle \equiv Q_{S^{\prime }S}$. Making use of the fractional
percentage coefficients by which the spin-state of a single particle can be
extracted from $\vartheta _{SM}$,\cite{bao1} the general expression of $%
Q_{S^{\prime }S}$ can be obtained as given in eq.(2) to (7) of ref.\cite%
{bao2}. It turns out that $Q_{S^{\prime }S}$ can be nonzero only if $%
S^{\prime }=S$\ or $S\pm 2$.

For Rb with $c_{2}<0$, the unperturbed g.s. in the spin-space is $\vartheta
_{NM}$. Due to the limited choices of $S$ and $S^{\prime }$, the
perturbative state up to the second order is
\begin{equation}
\Theta _{gs}\approx \vartheta _{NM}+\beta ^{\prime }\vartheta
_{N-2,M}+\beta \frac{Q_{N-4,N-2}}{Xc_{2}(4N-6)}\vartheta _{N-4,M}
\label{5}
\end{equation}%
where $\beta ^{\prime }=\beta (1+\frac{Q_{N-2,N-2}-Q_{N,N}}{Xc_{2}(2N-1)})$,
$\beta =\frac{Q_{N-2,N}}{Xc_{2}(2N-1)}$. From the general expression of $%
Q_{S^{\prime }S}$,
\begin{equation}
Q_{N,N}=q(M^{2}+N^{2}-N)/(2N-1),  \label{6}
\end{equation}%
\begin{equation}
Q_{N-2,N}=\frac{q}{2N-1}\left[\frac{2((N-1)^{2}-M^{2})(N^{2}-M^{2})}{(2N-3)}%
\right]^{\frac{1}{2}}  \label{7}
\end{equation}%
\begin{eqnarray}
&&Q_{N-2,N-2}  \notag \\
&=&\frac{q[M^{2}(2N+3)+2N^{3}-7N^{2}+5N-2]}{(2N-1)(2N-5)}
\label{8}
\end{eqnarray}
and $Q_{N-4,N}=0$.

On the other hand, for any total spin-state, the operator $\Sigma
_{i}f_{iz}^{2}$ is equivalent to $N-\overset{\wedge }{N_{0}}$, where $%
\overset{\wedge }{N_{0}}$\ is the operator for the number of $\mu =0$
particles. Thus we have
\begin{equation}
\frac{\langle \Theta _{gs}|H_{1}^{\prime }|\Theta _{gs}\rangle }{q\langle
\Theta _{gs}|\Theta _{gs}\rangle }=N-\frac{\langle \Theta _{gs}|\overset{%
\wedge }{N_{0}}|\Theta _{gs}\rangle }{\langle \Theta _{gs}|\Theta
_{gs}\rangle }\equiv N-N\bar{\rho}_{0}  \label{9}
\end{equation}%
Inserting eq.(\ref{5}) into (\ref{9}), and making use of eq.(\ref{6}) to (%
\ref{8}), we have the hyperfine density with $\mu =0$\ as
\begin{eqnarray}
\bar{\rho}_{0}&\approx& (N^{2}-M^{2})/[N(2N-1)] \\
&~& -\frac{1}{qN}(2\beta ^{\prime }Q_{N-2,N}+(\beta ^{\prime
})^{2}(Q_{N-2,N-2}-Q_{NN})) \notag   \label{10}
\end{eqnarray}
where the terms higher than $q^{2}$ are excluded. The first term is the $%
q\rightarrow 0$ limit of $\bar{\rho}_{0}$ for Rb and is denoted as $(\bar{%
\rho}_{0})_{q=0}$, which has been found before as given in eq.(18) of the
ref.\cite{bao2}. Thus eq.(\ref{10}) is a generalization of the previous
finding for magnetic field zero to nonzero. It turns out that $%
(N^{2}-M^{2})/[N(2N-1)]=(C_{1,0;N-1,M}^{N,M})^{2}$, where the Clebsch-Gordan
coefficient is introduced. This is because, when $q=0$, the total spin of
the g.s. $S_{g}=N$. Thus, each particle must couple to a $(N-1)-$body system
with total spin $N-1$. Thus the probability of each single particle in $\mu
=0$\ is given by the above square.

It is notable that $(\bar{\rho}_{0})_{q=0}$ is irrelevant to dynamics but
depends decisively on $M$. In particular, $(\bar{\rho}_{0})_{q=0}$ is close
to 1/2 when $M$\ is small, and is close to zero when $M\approx N$. Note
that, for the frame upon it $M$\ is defined, $M\approx 0$ implies that $%
S_{g} $ is lying close to the X-Y plane and therefore nearly half of the
particles are in $\mu =0$. While the case $M\approx N$ implies that $S_{g}$
is lying close to the Z-axis and therefore the number of $\mu =0$ particles
should be small. Note that the second term at the right of eq.(\ref{10}) is
positive (because $\beta $ is negative when $c_{2}<0$). Thus, starting from $%
(\bar{\rho}_{0})_{q=0}$, $\bar{\rho}_{0}$ keeps increasing with $B$. This is
natural because the g.s. would do its best to increase the number of $\mu =0
$ particles so as to reduce the quadratic Zeeman energy.

For Na with $c_{2}>0$, the unperturbed g.s. is $\vartheta _{MM}$. The
related matrix elements of $H_{1}^{\prime }$\ are
\begin{equation}
Q_{MM}=q(2MN+2N+M)/(2M+3)  \label{11}
\end{equation}%
\begin{equation}
Q_{M+2,M}=\frac{2q}{2M+3}[\frac{(M+1)(N-M)(N+M+3)}{2M+5}]^{\frac{1}{2}}
\label{12}
\end{equation}%
\begin{eqnarray}
Q_{M+2,M+2}&=&\frac{q}{(2M+3)(2M+7)}  \\
&~& [M^{2}(4N+2)+M(10N-5)+10N-6] \notag \label{13}
\end{eqnarray}%
The perturbative state up to the second order is
\begin{equation}
\Theta _{gs}\approx \vartheta _{MM}+\gamma ^{\prime }\vartheta
_{M+2,M}-\gamma \frac{Q_{M+4,M+2}}{Xc_{2}(4M+10)}\vartheta _{M+4,M}
\label{14}
\end{equation}%
where $\gamma ^{\prime }=\gamma (1+\frac{Q_{M+2,M+2}-Q_{MM}}{Xc_{2}(2N-1)})$%
, $\gamma =-\frac{Q_{M+2,M}}{Xc_{2}(2M+3)}$.

Inserting eq.(\ref{14}) into eq.(\ref{9}), we have
\begin{eqnarray}
\bar{\rho}_{0}&\approx& (N-M)/[N(2M+3)]-\frac{1}{qN}(2\gamma ^{\prime
}Q_{M+2,M}  \notag \\
&~&+(\gamma ^{\prime })^{2}(Q_{M+2,M+2}-Q_{MM}))  \label{15}
\end{eqnarray}%
where the terms higher than $q^{2}$ are excluded.. The first term is $(\bar{%
\rho}_{0})_{q=0}$ for Na. This form was first found in 2000 in the ref.\cite%
{tlh2,mko} and also in \cite{bao2} derived in a different way later. Thus
eq.(\ref{15}) is a generalization of the previous finding from $B$ zero to
nonzero. When $M=0$ the g.s. is in a pure polar phase and every particle is
in a singlet pair. In this case the first term becomes 1/3 as stated in ref.%
\cite{sta,bao2} and in the Theorem I of ref.\cite{tasa13}. Since the square
of the Clebsch-Gordan coefficient for the singlet pair $(C_{1,\mu ;1,-\mu
}^{00})^{2}=1/3$, the probability of the particle in $\mu $ is therefore
1/3. This is the physical background of this value.\ When $M$ increases even
a little from zero, the first term at the right of eq.(\ref{15}) will
decrease remarkably from 1/3 (say, when $M$ increases from 0 to $2$,\ $(\bar{%
\rho}_{0})_{q=0}$ decreases from 1/3 to $\sim 1/7$). This remarkable
decrease arises from the factor $2M+3$\ in the denominator. It implies that
the appearance of a few polarized particles among the big group of singlet
pairs could cause serious effect. Thus, for the case with $c_{2}>0$, the
sensitivity against $M$\ when $M$\ is small is notable. Starting from $(\bar{%
\rho}_{0})_{q=0}$, $\bar{\rho}_{0}$ keeps increasing with $B$ (because $%
\gamma $\ is negative) as in the previous case.

\subsubsection{4, Asymptotic behavior of the hyperfine density when $%
B\rightarrow \infty $}

When $B\rightarrow \infty $, $H_{0}^{\prime }$, rather than $H_{1}^{\prime }$%
, can be treated as a perturbation. Since the Fock-state $%
|N_{1},N_{0},N_{-1}\rangle $, in which $N_{\mu }$ particles are in the $\mu $%
-component, is an eigenstate of $H_{1}^{\prime }$, the perturbative
states can be expanded in terms of them. Since $N_{\pm
1}=(N-N_{0}\pm M)/2$, the Fock-state can be simply denoted as
$|N_{0}\rangle $ when $M$\ is fixed. Let $\langle N_{0}^{\prime
}|H_{0}^{\prime }|N_{0}\rangle \equiv P_{N_{0}^{\prime }N_{0}}$.
Making use of the formula, say, given in eq.(A5) of ref.\cite{ml},
we have
\begin{equation}
P_{N_{0}N_{0}}=\frac{Xc_{2}}{2}(M^{2}+N+N_{0}+2NN_{0}-2N_{0}^{2})
\label{16}
\end{equation}%
\begin{equation}
P_{N_{0}-2,N_{0}}=\frac{Xc_{2}}{2}[N_{0}(N_{0}-1)((N-N_{0}+2)^2-M^2)]^{\frac{%
1}{2}}  \label{17}
\end{equation}%
\begin{equation}
P_{N_{0}+2,N_{0}}=\frac{Xc_{2}}{2}[(N_{0}+1)(N_{0}+2)((N-N_{0})^2-M^2)]^{%
\frac{1}{2}}  \label{18}
\end{equation}
Otherwise, $P_{N_{0}^{\prime }N_{0}}=0$.

Note that, when $c_{2}\rightarrow 0$, the number of $\mu =0$\ particles in
the g.s. will be maximized so as to minimize the Zeeman energy. Thus, under
the conservation of $M$, the leading term of $\Theta _{gs}$ should be $%
|N-M\rangle $. Accordingly, the perturbative state up to the second order is
\begin{eqnarray}
\Theta _{gs}\approx &&|N-M\rangle +\delta^{\prime }|N-M-2\rangle  \notag \\
&+&\delta \frac{P_{N-M-4,N-M-2}}{4q}|N-M-4\rangle   \label{19}
\end{eqnarray}%
where $\delta ^{\prime }=\delta (1+\frac{P_{N-M-2,N-M-2}-P_{N-M,N-M}}{2q})$,
$\delta =\frac{P_{N-M-2,N-M}}{2q}$.

Inserting $\Theta _{gs}$\ into eq.(\ref{9}), we have
\begin{equation}
\bar{\rho}_{0}\approx \frac{N-M}{N}-\frac{2}{N}(\delta ^{\prime
})^{2} \label{20}
\end{equation}%
where higher order terms have been neglected. This formula holds for both Rb
and Na. Since the second term at the right of eq.(\ref{20}) is negative, the
first term appears as the upper bound.

\subsubsection{5, Applicability of the asymptotic expressions for c$_{2}<0$}

In order to clarify the applicability of the asymptotic expression, we have
to find out the exact solutions of $H^{\prime }$. It is reminded that the
set of eigenstates of $H_{0}^{\prime }$, $\{\vartheta _{SM}\}$, is complete
for all the symmetric spin-states, therefore they can serve as the basis
functions for the diagonalization of $H^{\prime }$ in the spin-space. The
related matrix element $\langle \vartheta _{S^{\prime }M}|H^{\prime
}|\vartheta _{SM}\rangle =\delta _{S^{\prime }S}\frac{Xc_{2}}{2}%
S(S+1)+Q_{S^{\prime }S}$, where $Q_{S^{\prime }S}$ is given in the ref.\cite%
{bao2}, and $X\equiv \int d\mathbf{r}|\phi |^{4}$ depends on the
interaction and the trap. To obtain numerical results,
$U(\mathbf{r})=\frac{1}{2}m\omega ^{2}r^{2}$ is assumed. Note that
$X$\ does not appear in the leading terms of all the asymptotic
expressions, and we consider only the cases with a large $N$. Thus,
it is reasonable to evaluate $X$ under the Thomas-Fermi
approximation (TFA). For details, we refer the reader to
ref.\cite{TF}. After the diagonalization one can extract
$\bar{\rho}_{0} $ from the eigenstates of $H^{\prime }$ as given in
eq.(9) of ref.\cite{bao2}. Numerical results of $\bar{\rho}_{0}$
from the asymptotic expressions and from the diagonalization are
compared below.

It turns out that, when $q=0$ and $q=\infty $, the resultant $\bar{\rho}_{0}$
from the perturbative approach and from the exact diagonalization of $%
H^{\prime }$\ are identical. Furthermore, we always have $(\bar{\rho}%
_{0})_{q\rightarrow \infty }\geq \bar{\rho}_{0}\geq $ $(\bar{\rho}%
_{0})_{q=0} $\ (where the equality holds only if $M=N$). Thus the variation
of $\bar{\rho}_{0}$\ versus $q$\ is strictly restricted in a domain which
can be known in advance, and $\bar{\rho}_{0}$ keeps increasing with $q$
inside the domain. The increasing arises because the g.s. would like to have
more $\mu =0$\ particles to reduce the quadratic Zeeman energy. We introduce
a ratio $\alpha $\ so that $M=\alpha N$. For Rb, we know from eq. (\ref{10})
and (\ref{20}) that $(\bar{\rho}_{0})_{q=0}\approx \frac{1}{2}(1-\alpha
^{2}) $\ and $(\bar{\rho}_{0})_{q\rightarrow \infty }=1-\alpha $. In what
follows, $\alpha $ is given at 1/2 and 0.001, respectively. Accordingly, the
magnetization is half and nearly zero. The case $\alpha \simeq 1$
(corresponding to a nearly full magnetization) is trivial because both the
lower and upper bounds are close to zero, and nearly no $\mu =0$\ particles
will emerge.

Examples of the variation of the densities $\bar{\rho}_{0}$ versus
$B$ with $N=10000$ and $\alpha =1/2$ (0.001) is shown in Fig.1
(Fig.2). Accordingly, $\bar{\rho}_{0}$ is increasing from 3/8 to 1/2
and from 1/2 to 0.999, respectively. To measure the deviation
between the exact and asymptotic results, we define $x_{i}=|(\bar{\rho}_{0})_{exact}-(\bar{\rho}%
_{0})_{asym,i}|$ where $(\bar{\rho}_{0})_{exact}$ denotes that the density
is from the exact diagonalization of $H^{\prime }$, while $(\bar{\rho}%
_{0})_{asym,i}$ is from the asymptotic expression eq.(\ref{10}) (if
$i=1$) or from eq.(\ref{20}) (if $i=2$). To give a quantitative
description, we define $B_{1}$ at which $x_{1}=0.01$, and $B_{2}$ at
which $x_{2}=0.01$. We found that, when $B<B_{1}$ ($>B_{2}$),
$x_{1}$ ($x_{2}$) is even smaller
than 0.01. Thus, we can say that the effective range of $B$, in which eq.(%
\ref{10}) is very close to be exact, is [$0,B_{1}$], and the effective range
for eq.(\ref{20}) is [$B_{2},\infty $].

The case with a half-magnetization ($\alpha =1/2$) is shown in Fig.1. In
Fig.1a $(B_{1},B_{2})=(343mG$, $361mG)$. Thus, the effective range for $(%
\bar{\rho}_{0})_{asym,1}$ and the effective range for $(\bar{\rho}%
_{0})_{asym,2}$ together cover nearly the whole range of $B$. It is
emphasized that both the lower and upper bounds depend only on $\alpha $ but
not on $N$\ and/or $\omega $. Therefore, when $N$\ and/or $\omega $\ are
changed, the variation of the curves is limited because they are fixed at
their two ends and they must keep increasing with $B$. Nonetheless, $B_{1}$
and $B_{2}$ will therefore be changed. For an example, the $\omega $\ in 1a
is two times the $\omega $\ in 1b. Accordingly, the curves in 1a as a whole
shift to the right. Numerical examples of $(B_{1},B_{2})$ with different $N$
and $\omega $ are given in Table I.

\begin{table}[pb]
\caption{$B_{1}$ and $B_{2}$ in $mG$ for Rb with $\protect\alpha \equiv
M/N=1/2$. Two cases of $N$ and three cases of $\protect\omega $ are
considered, where $\protect\omega _{o}=300\times 2\protect\pi /s$. }
\label{table.1}
\begin{center}
\begin{tabular}{llll}
&  & N=10000 & N=20000 \\
$\omega _{o}/2$ &  & 226, 238 & 260, 274 \\
$\omega _{o}$ &  & 343, 361 & 393, 414 \\
$2\omega _{o}$ &  & 520, 548 & 596, 629%
\end{tabular}%
\end{center}
\end{table}

We know from the Table that $B_{2}$ is close to $B_{1}$ in general, and
therefore the asymptotic expressions are nearly valid in the whole range of $%
B$. They will both be larger when $N$\ and/or $\omega $\ are larger, or vice
versa. Nonetheless, they are more sensitive to $\omega $\ rather than $N$.
When $\omega $\ is larger, the spin-texture of the g.s. would have a
stronger ability to resist the field. This causes a shift of the curves to
the right.

The case with a nearly zero magnetization ($\alpha =0.001$) is shown in
Fig.2. It was found in 2a (2b) that both $B_{1}$ and $B_{2}$ are very close
to a critical point $B_{crit}=446mG$ (292$mG$) at which $(\bar{\rho}%
_{0})_{exact}$ undergoes a sharp change. The change is so sharp that the
second order derivative of $(\bar{\rho}_{0})_{exact}$ against $B$\ tends to $%
\infty $\ when $N\rightarrow \infty $. Obviously, this implies a phase
transition. It turns out that the two curves $(\bar{\rho}_{0})_{asym,i}$ and
$(\bar{\rho}_{0})_{asym,i}$ intercept when $B\sim B_{crit}$, and this is a
common feature for the g.s. with $c_{2}<0$ and with a nearly
zero-magnetization. Thus the effective ranges for the two asymptotic
expressions together cover the whole range of $B$, and the two expressions
together provide a perfect description of $\bar{\rho}_{0}$. Numerical
examples of $B_{crit}$ are given in Table II.

\begin{table}[pb]
\caption{$B_{crit}$ in $mG$ for Rb with $\protect\alpha =0.001$. In this
case $B_{1}\approx B_{2}\approx B_{crit}$. $\protect\omega _{o}$ is referred
to Table I.}
\label{table.2}
\begin{center}
\begin{tabular}{llll}
&  & N=10000 & N=20000 \\
$\omega _{o}/2$ &  & 292 & 337 \\
$\omega _{o}$ &  & 446 & 514 \\
$2\omega _{o}$ &  & 678 & 778%
\end{tabular}%
\end{center}
\end{table}

Similar to the previous case, a larger $N$\ and/or a larger $\omega $\ lead
to a larger $B_{crit}$, and the vice versa. When $B>B_{crit}$, the curve of $%
(\bar{\rho}_{0})_{exact}$ becomes horizontal. It implies that, after the
phase transition, the system arrives at its eventual status (this status is
described by the Fock-state $|M,N-M,0\rangle $, where $N-M$ particles have $%
\mu =0$ while the rest have $\mu =1$). In this status the number of $\mu =0$
particles has been maximized, and therefore no more change is allowed when $%
B $ increases further. Note that the sharp change will become ambiguous when
$M $\ is larger. Comparing 2b with 2a, it is clear that $B_{crit}$ will
shift to the right when $\omega $\ increases.

\subsubsection{6, Applicability of the asymptotic expressions for c$_{2}>0$%
}

Let us predict two features of spin-1 condensates with $c_{2}>0$.

(i) When $B=0$, the total spin-state of the g.s. has $S=M$\ and is denoted
as $\vartheta _{MM}\propto \overset{\symbol{126}}{\mathcal{S}}\chi
_{1}^{M}(\chi \chi )_{0}^{(N-M)/2}$, where $\chi _{1}$ denotes the
spin-state of a single particle with $\mu =1$, $(\chi \chi )_{0}$ denotes a
singlet pair, and $\overset{\symbol{126}}{\mathcal{S}}$ is a symmetrizer. If
there are another forms, they are identical due to the uniqueness of the
eigenstate.\cite{ka} Thus, the g.s. is a mixture of a group of polarized
particles together with a group of singlet pairs. Obviously, the $M$
particles in the first group are stable against $B$, while the $N-M$
particles in the second group are not. Note that every singlet pair is
situated under the same environment, thus they have similar ability to
resist $B$. It is possible that all the pairs might begin to be broken when $%
B$ increases and exceeds a certain value, and therefore a sharp change of
the spin-texture will occur.

(ii) In general, the stability of the g.s. depends on the gap (the energy
difference between the g.s. and the first excited state). When $B=0$ the
g.s. has $S=M$\ while the first excited state has $S=M+2$, therefore the\
gap is $\propto (4M+6)$ for $c_{2}>0$, whereas this factor would be $(4N-2)$
for $c_{2}<0$. Therefore, the gap for Na is much smaller than that for Rb
when $\alpha $\ is small. In this case, the g.s. of Na is highly unstable
and the feature of the system will depend on $\alpha $\ sensitively. In
other words, the singlet pairs will become more fragile when $\alpha
\rightarrow 0$.

We found from eq.(\ref{15}) and eq.(\ref{20}) that the lower and
upper bounds are$\ (\bar{\rho}_{0})_{q=0}=(1-\alpha )/(2\alpha N+3)$\ and $(%
\bar{\rho}_{0})_{q\rightarrow \infty }=1-\alpha $, respectively, and they
are identical to the numerical values from the exact diagonalization of $%
H^{\prime }$. In particular, when $\alpha =0$, $(\bar{\rho}_{0})_{q=0}=1/3$\
and $(\bar{\rho}_{0})_{q=\infty }=1$. Note that, when $\alpha =0$, every
particle is in a singlet pair. Recalled that the Clebsch-Gordan coefficient $%
C_{1,\mu ;1,-\mu }^{0,0}=(-1)^{1-\mu }1/\sqrt{3}$. Therefore, when a
particle is in a singlet pair, the probability in $\mu $ is 1/3. This leads
to $(\bar{\rho}_{0})_{q=0}=1/3$. On the other hand, $(\bar{\rho}%
_{0})_{q=\infty }=1$ implies obviously that all particles are in the $\mu =0$
component. This results from the minimization of the quadratic Zeeman energy.

For any $\alpha $, we have $(\bar{\rho}_{0})_{q=0}\leq \bar{\rho}_{0}\leq (%
\bar{\rho}_{0})_{q\rightarrow \infty }$, and $\bar{\rho}_{0}$ keeps
increasing in between as before. Note that the $(\bar{\rho}%
_{0})_{q\rightarrow \infty }$ of the two species Rb and Na are the same, but
the $(\bar{\rho}_{0})_{q=0}$ of Na is always lower than that of Rb. Thus,
when $q$ varies, the $\bar{\rho}_{0}$ of Na varies in a broader domain. Note
that the $\alpha -$sensitivity of $(\bar{\rho}_{0})_{q=0}$ is embodied in
the factor $2\alpha N$\ which appears in the denominator (Say, when $N=10000$%
, and $\alpha =0$ and $0.001$, respectively, $(\bar{\rho}_{0})_{q=0}=1/3$
and $1/23$. Thus a small change in $\alpha $ leads to a big change in $(\bar{%
\rho}_{0})_{q=0}$).

For the case of half-magnetization ($\alpha =1/2$), the variation of $\bar{%
\rho}_{0}$ versus $B$\ is shown in Fig.3 where $N=10000$\ and $\omega
=\omega _{o}\equiv 300\times 2\pi /s$. There is a sudden uprising in $(\bar{%
\rho}_{0})_{exact}$\ as mentioned in point (i) taking place when $%
B=B_{crit}=114mG$. It turns out that $(\bar{\rho}_{0})_{asym,1}$ is
identical or extremely close to $(\bar{\rho}_{0})_{exact}$ when $B\leq
B_{crit}$, but deviates from $(\bar{\rho}_{0})_{exact}$\ rapidly when $%
B>B_{crit}$. Therefore, in this case, we have $B_{1}\approx B_{crit}$\ and
therefore the effective range for $(\bar{\rho}_{0})_{asym,1}\ $is $(0,
B_{crit})$. Numerical examples of $B_{crit}$ and $B_{2}$ (the latter is much
larger) are shown in Table III.

\begin{table}[pb]
\caption{$B_{1}\approx B_{crit}$ and $B_{2}$\ in $mG$ for Na with $\protect%
\alpha =1/2$. Refer to Table I.}
\label{table.3}
\begin{center}
\begin{tabular}{llll}
&  & N=10000 & N=20000 \\
$\omega _{o}/2$ &  & 77, 226 & 90, 259 \\
$\omega _{o}$ &  & 118, 342 & 136, 393 \\
$2\omega _{o}$ &  & 176, 519 & 204, 595%
\end{tabular}%
\end{center}
\end{table}

Thus, both $B_{1}$ and $B_{2}$ will be larger when $N$\ and/or $\omega $\
are larger as before, or the vice versa. In fact, $B_{crit}$ measures the
ability of the singlet pairs to keep themselves unbroken against $B$. It is
obvious that this ability will become stronger when the trap is stronger ($%
\omega $ is larger). It is interesting to see that this ability will also
become a little stronger when $N$\ increases.

\textit{Recalled that the }$B_{crit}$\textit{\ for Rb marks the maximization
of the }$\mu =0$\textit{\ particles. Now, the }$B_{crit}$\textit{\ for Na
marks the solidity of the singlet pairs.} This explains why the curve is
horizontal when $B>B_{crit}$ for Rb, and when $B<B_{crit}$ for Na. In the
latter case the pairs are kept unbroken and the spin-texture remains
unchanged.

The case of nearly zero-magnetization ($\alpha =0.001$) is shown in Fig.4.
It is shown in Fig.4a (for $B\leq 1mG$) that $(\bar{\rho}_{0})_{exact}$
increases very fast when $B$\ is ranged from $0.2mG$ to $1mG$. One can
define a critical strength $B_{crit}$ at which the second order derivative
of $(\bar{\rho}_{0})_{exact}$\ against $B$\ arrives at its maximum. In
Fig.4a $B_{crit}=0.26mG$. It is further found that the effective range of $B$
for $(\bar{\rho}_{0})_{asym,1}$ to be applicable is from 0 to $B_{1}=0.29mG$%
. Thus $B_{1}$ is again very close to $B_{crit}$. Once $B>B_{crit}$,
although both $(\bar{\rho}_{0})_{asym,1}$ and $(\bar{\rho}_{0})_{exact}$
increase sharply, they deviate more and more from each other. For the case
that $\alpha $\ is very small, the effective range for $(\bar{\rho}%
_{0})_{asym,1}$ is very narrow. As an example, if $\alpha $\ is further
reduced from 0.001 to zero, $B_{1}$ would be reduced from 0.29mG to 0.04mG,
and $B_{crit}$\ is reduced from 0.26mG to zero. Thus, the singlet pairs will
become very fragile when $\alpha \rightarrow 0$ as predicted in (ii).

How $B_{1}$ and $B_{2}$\ are affected by $N$ and $\omega $ is shown
in Table IV.

Table IV, $B_{1}$ and $B_{2}$\ in $mG$ for Na with $\alpha =0.001$. Refer to
Table I.

\begin{table}[pb]
\caption{$B_{1}$ and $B_{2}$\ in $mG$ for Na with $\protect\alpha =0.001$.
Refer to Table I.}
\label{table.4}
\begin{center}
\begin{tabular}{llll}
&  & 10000 & 20000 \\
$\omega _{o}/2$ &  & 0.17, 163 & 0.18, 187 \\
$\omega _{o}$ &  & 0.26, 248 & 0.27, 282 \\
$2\omega _{o}$ &  & 0.39, 375 & 0.41, 428%
\end{tabular}%
\end{center}
\end{table}

\subsubsection{7, Summary}

When $B$\ tends to zero and infinite, the asymptotic forms of the hyperfine
density $\bar{\rho}_{0}$ against $B$ denoted as $(\bar{\rho}%
_{0})_{asym,1}$ and $(\bar{\rho}_{0})_{asym,2}$, respectively, has
been obtained analytically based on the second order perturbation
theory. Numerical calculation via an exact diagonalization of
$H^{\prime }$\ for the density denoted as
$(\bar{\rho}_{0})_{exact}$\ has been performed to clarify
the effective ranges of $B$ wherein the asymptotic forms $(\bar{\rho}%
_{0})_{asym,i}$\ are applicable. The two ranges for $i=1$ and $i=2$,
respectively, are specified as $(0,B_{1}$) and $(B_{2},\infty )$. It turns
out that the two limits $(\bar{\rho}_{0})_{q=0}$ and $(\bar{\rho}%
_{0})_{q=\infty }$ given by the analytical expressions are identical to
those from the exact diagonalization of $H^{\prime }$. They together provide
the lower and upper bounds for $\bar{\rho}_{0}$, and $\bar{\rho}_{0}$ is
monotonously increasing with $q$ between them. Thus, in any case, we can
have a rough impression on $\bar{\rho}_{0}$.

For Rb, both $B_{1}$ and $B_{2}$ are in the order of $10^{2}mG$, and they
are close to each other when the parameters vary in a broad domain that are
accessed frequently in experiments. Thus, $\bar{\rho}_{0}$ can be accurately
known when $B$\ lies inside one of the ranges, or can be roughly known when $%
B$\ lies between $B_{1}$ and $B_{2}$. In particular, when $\alpha \equiv M/N$%
\ is small, (i) \textit{A critical strength }$B_{crit}$\textit{\ is found
which marks the realization of the eventual status, in which the number of }$%
\mu =0$\textit{\ particles is maximized.} (ii) $B_{1}$, $B_{2}$, and $%
B_{crit}$ are close to each other, and therefore the two asymptotic forms
together cover nearly the whole range.

\textit{For Na, a critical strength }$B_{crit}$\textit{\ is also found but
has another implication, it marks the sudden breaking of the singlet pairs. }%
At a weak field $\bar{\rho}_{0}$ is highly sensitive to $\alpha $\ when $%
\alpha $\ is small. Note that, when $\alpha $\ increases from zero, a few
polarized particles will emerge among the numerous singlet pairs and lead to
a remarkable increase of $B_{crit}$. It implies that the singlet pairs will
become more solid thereby. Thus the emergence of a few specific particles
can cause serious effect, similar to the serious effect of a few impurity
appearing in well organized crystal structure. The underlying physics
deserves to be further studied.

In general, when the spin-texture undergoes a sharp change (this appears as
a phase transition when $N\rightarrow \infty $), $(\bar{\rho}_{0})_{exact}$
is able to describe this change. However, the perturbation theory fails to
describe such a sharp change. This is the main shortcoming of the
perturbation theory and is the reason that $(\bar{\rho}_{0})_{asym,i}$ will
begin to deviate sharply from $(\bar{\rho}_{0})_{exact}$ when $q$\ is close
to the critical points.

\begin{acknowledgments}
Supported by the National Natural Science Foundation of China (NNSFC) under
Grant No.11372122, the Open Project Program of State Key Laboratory of
Theoretical Physics, Institute of Theoretical Physics, Chinese Academy of
Sciences, China
\end{acknowledgments}

\newpage

\begin{figure}[tbp]
\begin{center}
\resizebox{0.95\columnwidth}{!}{\includegraphics{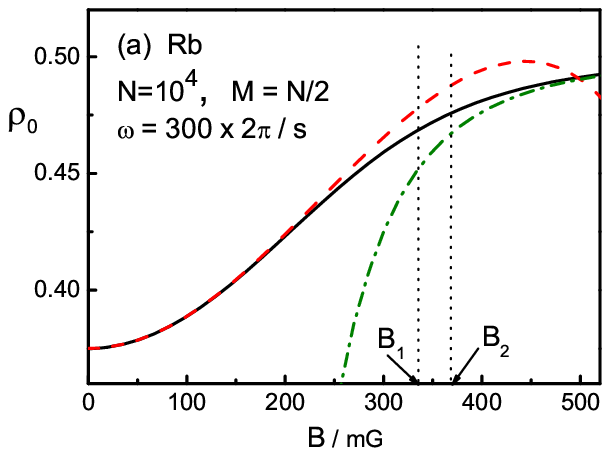} }
\end{center}
\end{figure}

\begin{figure}[tbp]
\begin{center}
\resizebox{0.95\columnwidth}{!}{\includegraphics{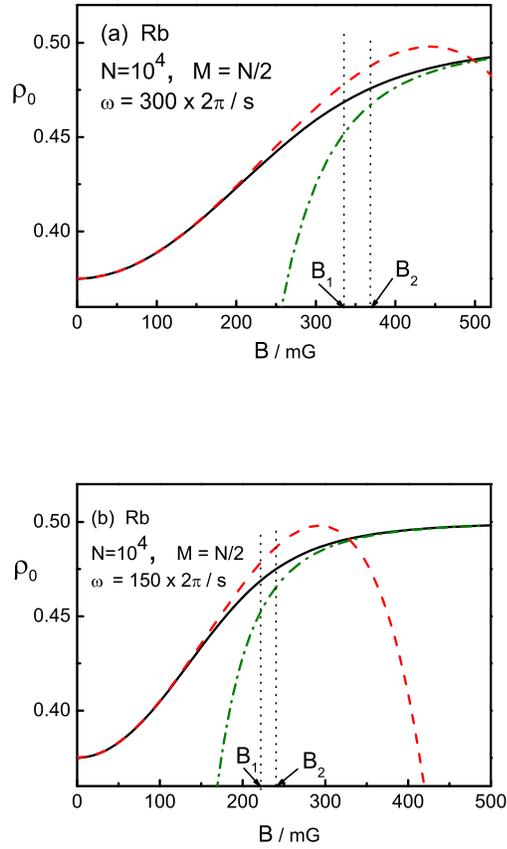} }
\end{center}
\caption{(Color online) The three densities $(\bar{\protect\rho}_{0})_{exact}
$ (solid), $(\bar{\protect\rho}_{0})_{asym,1}$(dash), and $(\bar{\protect\rho%
}_{0})_{asym,2}$(dash-dot) of Rb versus $B/mG$. $M=N/2$. The trap is assumed
to be $\frac{1}{2}m\protect\omega ^{2}r^{2}$. $(\bar{\protect\rho}%
_{0})_{asym,1}$ is effective (close to $(\bar{\protect\rho}_{0})_{exact}$)
when $B$\ lies in $(0,B_{1})$, while $(\bar{\protect\rho}_{0})_{asym,2}$ is
effective in $(B_{2},\infty )$. $B_{1}$ and $B_{2}$ are marked by two
vertical dotted lines. Note that the upper panel has a larger $\protect%
\omega $, and accordingly the curves shift to the right. It implies that,
for a stronger trap, a stronger $B$ is needed to raise up the number of $%
\protect\mu =0$ particles.}
\label{fig:1}
\end{figure}

\begin{figure}[tbp]
\begin{center}
\resizebox{0.95\columnwidth}{!}{\includegraphics{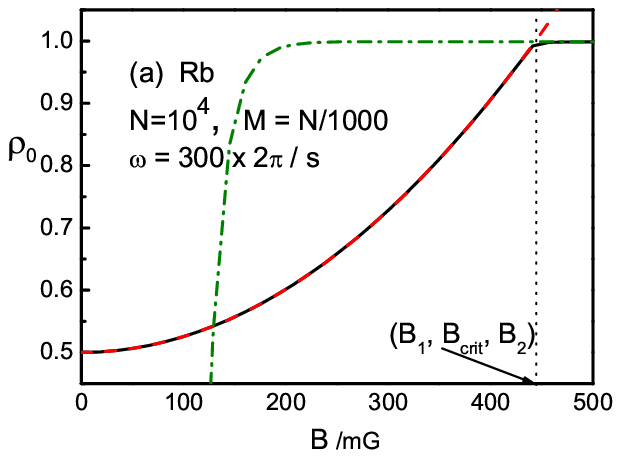} }
\end{center}
\end{figure}

\begin{figure}[tbp]
\begin{center}
\resizebox{0.95\columnwidth}{!}{\includegraphics{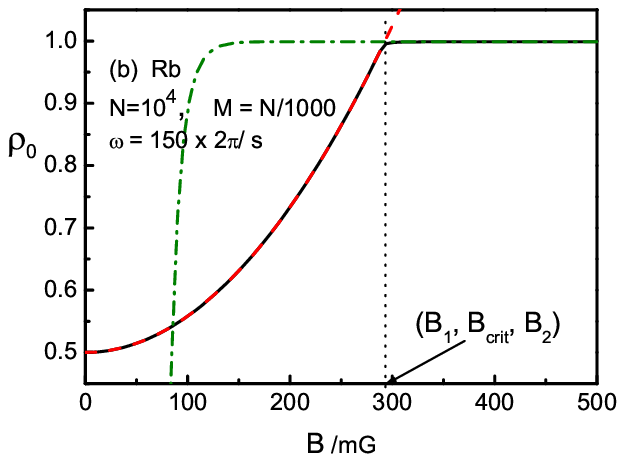} }
\end{center}
\caption{(Color online) The same as Fig.1 but with $M=N/1000$. There is a
sharp change in $(\bar{\protect\rho}_{0})_{exact}$\ at $B_{crit}$ marked by
a vertical dotted line. $B_{1}\approx $ $B_{crit}\approx B_{2}$\ is found.}
\label{fig:2}
\end{figure}

\begin{figure}[tbp]
\begin{center}
\resizebox{0.95\columnwidth}{!}{\includegraphics{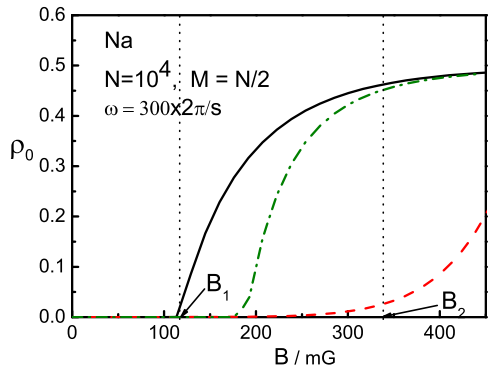} }
\end{center}
\caption{(Color online) The three densities of Na versus $B/mG$. Refer to
Fig.1.}
\label{fig:3}
\end{figure}

\begin{figure}[tbp]
\begin{center}
\resizebox{0.95\columnwidth}{!}{\includegraphics{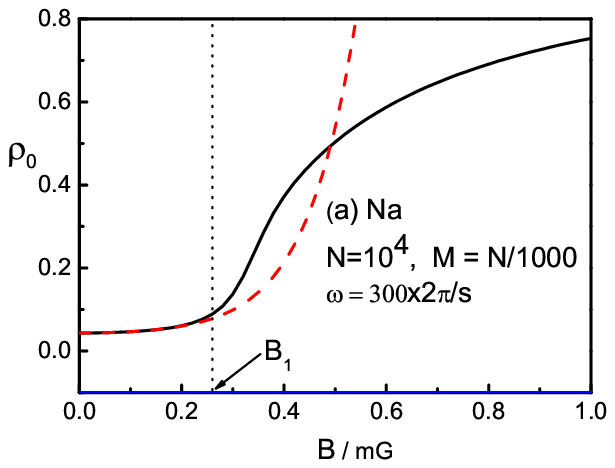} }
\end{center}
\end{figure}

\begin{figure}[tbp]
\begin{center}
\resizebox{0.95\columnwidth}{!}{\includegraphics{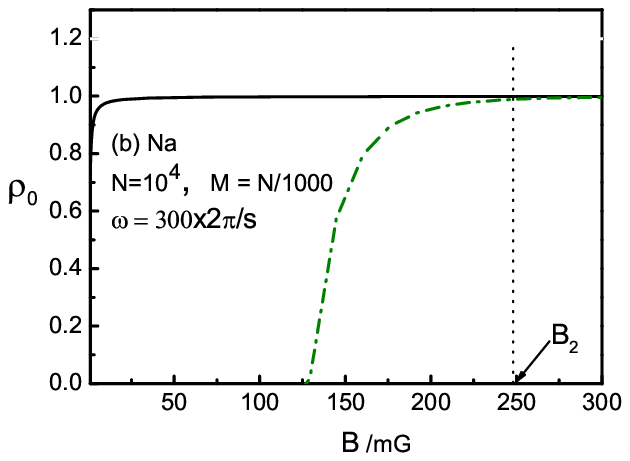} }
\end{center}
\caption{(Color online) The same as Fig.3 but with $M=N/1000$. The range of $%
B$\ is $0\rightarrow 1mG$ in 3a where $B_{1}$\ is marked, whereas $B$\ is $%
1\rightarrow 300mG$ in 4b where $B_{2}$\ is marked. $(\bar{\protect\rho}%
_{0})_{asym,2}$ does not appear in 4a, while $(\bar{\protect\rho}%
_{0})_{asym,1}$\ does not appear in 4b.}
\label{fig:4}
\end{figure}


\begin{thebibliography}{99}
\bibitem{TF} \ In our numerical calculation $U(\mathbf{r})=\frac{1}{2}%
m\omega ^{2}r^{2}$ is assumed. $\hbar \omega $, $\lambda \equiv \sqrt{\hbar
/(m\omega )}$, and $Gauss$\ are used as units for energy, length, and $B$,
respectively. We have defined five quantities $\bar{c}_{0}$, $\bar{c}_{2}$, $%
\bar{\omega}$, $\bar{q}$ and $\bar{B}$. They are the dimensionless
correspondents of $c_{0}$ to $B$ (say $c_{0}=\bar{c}_{0}\hbar \omega \lambda
^{3}$, $B=\bar{B}G$, etc.). In realistic cases, their magnitudes are $\bar{c}%
_{0}=6.77\times 10^{-4}\sqrt{\bar{\omega}}$ ($2.49\times 10^{-3}\sqrt{\bar{%
\omega}}$), $\bar{c}_{2}=2.12\times 10^{-5}\sqrt{\bar{\omega}}$ ($%
-1.16\times 10^{-5}\sqrt{\bar{\omega}}$), and $\bar{q}=1745\bar{B}^{2}/\bar{%
\omega}$ ($452\bar{B}^{2}/\bar{\omega}$) for Na (Rb). The Thomas-Fermi
approximation leads to $\phi \simeq \sqrt{\frac{15}{8\pi r_{0}^{3}}(1-\frac{%
r^{2}}{r_{0}^{2}})}$ when $r\leq r_{0}$, or $\phi =0$ when $r>r_{0}$, and $%
r_{0}=(\frac{15N\bar{c}_{0}}{4\pi })^{1/5}$. Thereby, $X\simeq 0.3067(N\bar{c%
}_{0})^{-3/5}\lambda ^{-3}\equiv X_{\mathrm{TF}}$.
\end{thebibliography}
\end{document}